\documentclass[]{scrartcl}

\author{\large Urs A. T. Hofmann$^{1,2,+}$ \and \large Sergio Pérez-López$^{3,+}$ \and \large Héctor Estrada$^{1,2,*}$ \and \large Daniel Razansky$^{1,2,*}$}
\title{\huge High order pulse-echo (HOPE) ultrasound}
\date{\normalsize
    $^1$Institute for Biomedical Engineering, Department of Information Technology and Electrical Engineering, ETH Zurich, Switzerland\\%
    $^2$Institute of Pharmacology and Toxicology, Faculty of Medicine, University of Zurich, Switzerland\\
    $^3$Centro de Tecnologías Físicas, Universitat Politècnica de València, Spain\\
    ${+}$ These authors contrinuted equally to this work\\
    Corresponding authors: \href{hector.estrada@posteo.org}{\texttt{hector.estrada@posteo.org}}, \href{daniel.razansky@uzh.ch}{\texttt{daniel.razansky@uzh.ch}}\\[2ex]%
    \today
}

\usepackage{siunitx}
\usepackage{amsmath}
\usepackage{xcolor}
\usepackage{graphicx}
\usepackage[left=20mm, right=20mm]{geometry}

\usepackage{acronym}
\newacro{CNR}{contrast-to-noise ratio}
\newacro{HEK}{human embryonic kidney}
\newacro{HOPE}{high order pulse-echo}
\newacro{MAP}{maximum amplitude projection}
\newacro{NDE}{non-destructive evaluation}
\newacro{PBS}{phosphate buffer solution}
\newacro{PVDF}{polyvinylidene fluoride}
\newacro{SAM}{scanning acoustic microscopy}
\newacro{SNR}{signal-to-noise ratio}
\newacro{US}{ultrasound}

\usepackage[footnotesize,bf]{caption}
\setcapindent{0pt}
\usepackage{sidecap}

\usepackage[square,numbers,sort&compress]{natbib}
\usepackage{hyperref}
\hypersetup{
	citecolor=red!75!black,
	linkcolor=red!75!black,
	urlcolor=green!65!black,
	colorlinks=true
}

\begin{document}

	\maketitle

	\begin{abstract}
		Multiple reflections between transducer and imaged object can naturally occur in ultrasound imaging and other acoustic sensing applications such as sonar. The repeated interaction of the emitted wavefront with the imaged object is traditionally regarded an undesired reverberation artifact, often misinterpreted as fictitious acoustic boundaries.  Here we introduce high order reflection pulse-echo (HOPE)-ultrasound, a novel method that leverages high order reflections to improve on several aspects of conventional ultrasound imaging. HOPE is experimentally demonstrated to resolve sub-micrometer features by breaking through the Nyquist resolution limit. The major contrast enhancement of the high reflection orders allowed to reveal defects within materials invisible to conventional scanning acoustic microscopy. The technique is further shown to improve accuracy of frequency-dependent ultrasound attenuation measurements from biological tissues. HOPE ultrasound requires no additional hardware and is easy to implement, underscoring its potential to boost imaging performance in a broad range of biomedical imaging, non-destructive testing, and other acoustic sensing applications.
	\end{abstract}



\section{Introduction}

Pulse-echo \ac{US} imaging systems are used to investigate materials and biological samples at microscopic and macroscopic scales. Compared to other imaging technologies,  such as X-ray / computed tomography and magnetic resonance imaging, \ac{US}-based modalities offer several major advantages, including ease of use, affordability, and portability \cite{wells_ultrasonic_1999,deffieux_functional_2018,briggs_acoustic_2010}. 
Owing to its nonionizing and noninvasive nature, \ac{US} imaging has also found broad applicability in biology and medicine. The precise record of the time of flight of \ac{US} waves reflected at boundaries between acoustically mismatched materials allows rendering volumetric images from samples at high spatial resolution. Extensions to the conventional pulse-echo technique include speed of sound imaging, elastography, and Doppler imaging, methods that exploit complementary types of mechanical contrast further providing valuable functional information when imaging through living tissues \cite{duric_breast_2013,hong_microscale_2016,sigrist_ultrasound_2017,poelma_ultrasound_2017}. 

\begin{figure}[hbt]
	\centering
	\includegraphics[width=0.6\columnwidth]{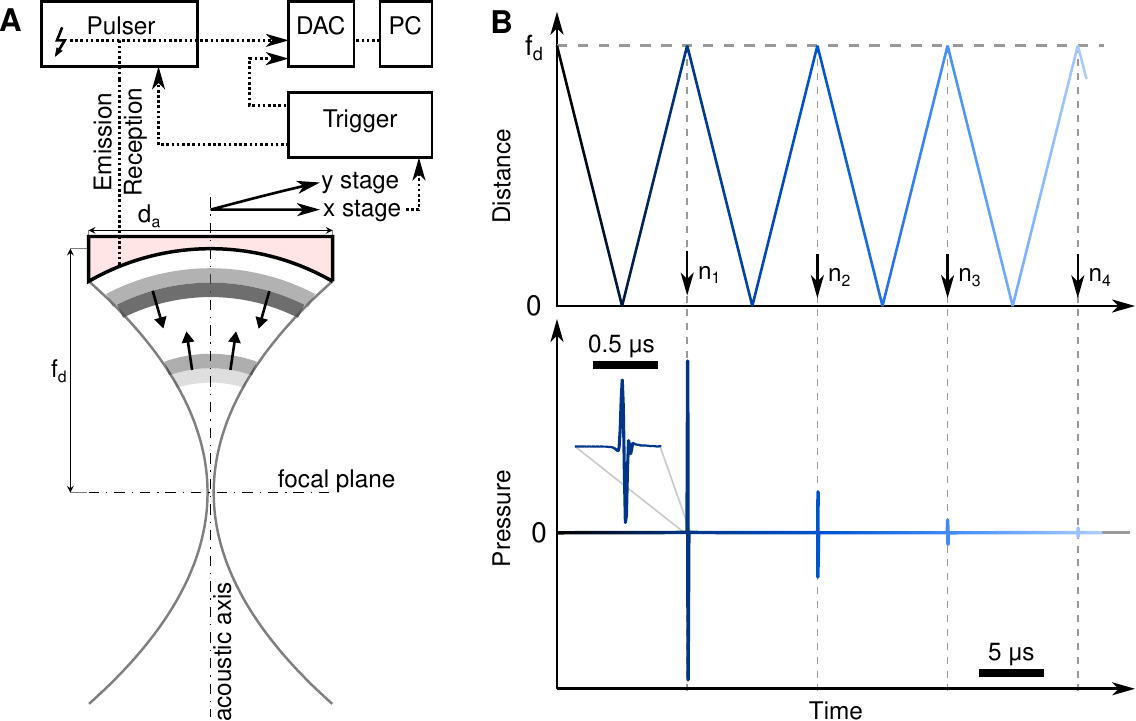}
	\caption{Experimental concept of \ac{HOPE} ultrasound. (A) In reflection mode \ac{SAM}, a single element US transducer is moved over the imaged volume by two perpendicularly arranged (X-Y) stages while emitting and receiving a reflected pulse at each grid position. The spherically shaped polymer-based element made of \ac{PVDF} focuses the emitted \ac{US} waveform onto the sample while detecting reflections over an ultrawideband frequency range. The trigger board automatically reads out the stage position and time while triggering the pulsing and data acquisition (DAC) for fast overfly 3D imaging. (B) Due to acoustic impedance mismatch between sensitive element and coupling medium, the wavefronts get reflected back and forth multiple times between acoustic boundaries in the sample and the transducer surface, leading to high order reflections with decaying signal amplitude.}
	\label{fig1}
\end{figure}

In addition to a broad range of biomedical applications chiefly relying on array-based imaging, \ac{SAM} is widely employed in material science and industrial applications for \ac{NDE} of materials and mechanical or electrical components, reducing production and maintenance time and cost \cite{briggs_acoustic_2010,estrada_hybrid_2014,errico_ultrafast_2015,winterroth_comparison_2011,zheng_spectral_2007,maev_acoustic_2008,kundu_advanced_2007}. 
In conventional reflection-mode \ac{SAM}, a piezoelectric US transducer is pulsed by a short high voltage signal, leading to mechanical strain followed by the emission of a high-frequency \ac{US} wave. The latter is directed towards or focused into the sample by proper shaping of the active piezoelectric element (Fig.~\ref{fig1}A) or, alternatively, by guiding the emitted wave through an acoustic lens. The emitted wave is subsequently reflected at boundaries between acoustically mismatched materials, traveling back to the piezoelectric element that converts mechanical strain into voltage \cite{maev_acoustic_2008}. 
By analyzing the received signal amplitude and extracting the temporal delay between emission and reception, both acoustic reflectivity and distance of an acoustic boundary can be calculated, under the assumption of a known speed of sound. By scanning the transducer over a two-dimensional plane, volumetric images can be rendered. In most cases, after being received by the \ac{US} transducer, the acoustic waves get partially reflected back towards the sample due to an impedance mismatch between the piezoelectric material and the surrounding coupling medium, resulting in a longer pattern of pulses (Fig.~\ref{fig1}B). Those high order reflections following the first pulse-echo signal, also known as reverberations or multiple reflections, are considered an undesired artifact due to their common misinterpretation as additional (fictious) acoustic boundaries positioned deeper in the imaged sample. Numerous methods were previously proposed to suppress or remove such artifacts in \ac{SAM} and other applications relying on reflection of mechanical waves in matter \cite{hassler_apparatus_nodate,robinson_artefacts_1966,suzuki_method_nodate-1}.

Here we introduce the \ac{HOPE} ultrasound method making use of the information accumulated in the \ac{US} wave during high order reflections. We observed up to four-fold depth resolution increase in experimental profilometry measurements with \ac{HOPE}. Features invisible to conventional pulse-echo imaging become clearly discernible in the high order reflections due to significant enhancement of their respective \ac{CNR}. Moreover, \ac{HOPE} ultrasound allows for more accurate self-validated \ac{US} attenuation measurements. We showcase performance of the technique for improved distance measurements, material defect detection, microchip characterization, and estimation of acoustic attenuation properties of biological and non-biological samples.


\section{Results}

At each point in the scanned region, a single excitation pulse is emitted by the spherically shaped \ac{PVDF} foil-based transducer, propagating and converging towards the focus (Fig.~\ref{fig1}A). The target partially reflects the wave back to the transducer, where it is detected ($n_1$, Fig.~\ref{fig1}B) and again partially reflected towards the target. The succession of multiple reflections ($n_1$, $n_2$ , ..., $n_i$) generates a pulse train with decaying amplitude (Fig.~\ref{fig1}B) in which every new reflection order is iteratively convolved in space and time with the transducer's impulse response and the acoustic properties of the target.

\begin{figure}[h!bt]
	\centering
	\includegraphics{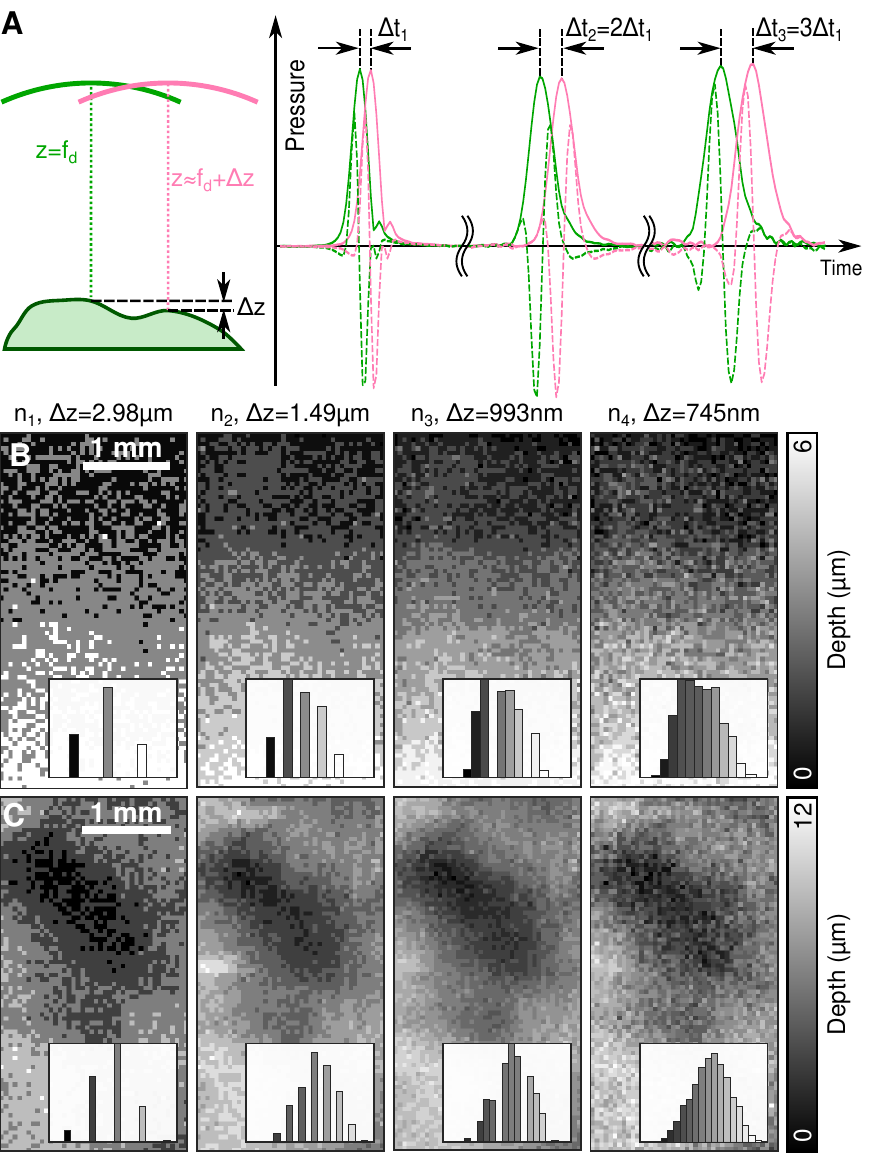}
	\caption{\ac{HOPE} improves the depth resolution beyond the quantization limit. (A) The difference in time-of-flight between two distinctively separated surface positions accumulates linearly with the reflection order, thus leading to an increase in depth resolution analogous to an increase in the sampling frequency. (B) Tilted flat glass and (C) Aluminum foil surface resolved using a succession of reflection orders acquired with a single scan. The left panel shows the first reflection (conventional pulse-echo) followed by higher order reflections on the right. The insets show the depth distribution.}
	\label{fig2}
\end{figure}

\paragraph{Accuracy of profilometry increases linearly with the reflection order}
Two acoustic boundaries located at different distances would result in time traces with peak locations shifted by a time delay corresponding to the separation between the boundaries (Fig.~\ref{fig2}A). The distance between the two acoustic boundaries can unambiguously be resolved if the difference in the time of arrival is larger than temporal sampling (discretization) during analog-to-digital conversion. The depth resolution, $\Delta z$, defined as the minimum resolvable axial distance, can therefore be expressed as a function of the reflection order $n$, the sampling frequency $f_s$, and the speed of sound in the coupling medium $c_0$, as
\begin{equation}
	\Delta z = \frac{c_0}{2 \cdot n \cdot f_s}
\end{equation}
which implies that the depth resolution increases linearly with the reflection order due to the time delay accumulating with each reflection (Fig.~\ref{fig2}A). This effect is only limited by the decreasing signal intensity and, therefore, \ac{SNR}.

A glass plate positioned under a slight angle against the scanning plane was imaged at a bidirectional step size of \SI{50}{\micro\meter}. At each position, an A-scan was acquired containing multiple high order reflections. At a constant sampling frequency of \SI{250}{\mega\hertz}, the smallest detectable runtime difference between two peaks of the \ac{US} wave corresponds to \SI{4}{\nano\second}, resulting in the axial spatial discretization of \SI{2.98}{\micro\meter} when considering a $c_0$ = \SI{1490}{\meter\per\second} speed of sound in the coupling medium. With increasing number of high order reflections, we could show a linear increase in detectable axial time-of-flight difference to allow for detection of sub-micrometer features down to \SI{745}{\nano\meter} in size with a \SI{30}{\mega\hertz} central frequency transducer, after detecting four reflections (Fig.~\ref{fig2}B).

Also, for non-even surfaces such as a \SI{12}{\micro\meter} thin aluminum foil (Fig.~\ref{fig2}C), \ac{HOPE} ultrasound overcomes the limitations of conventional pulse-echo \ac{US} by refining the depth resolution with measurements exceeding the quantization limit.

\paragraph{Contrast enhancement through high order reflections follows a geometric sequence}

\begin{figure}[h!bt]
	\centering
	\includegraphics[width=0.5\columnwidth]{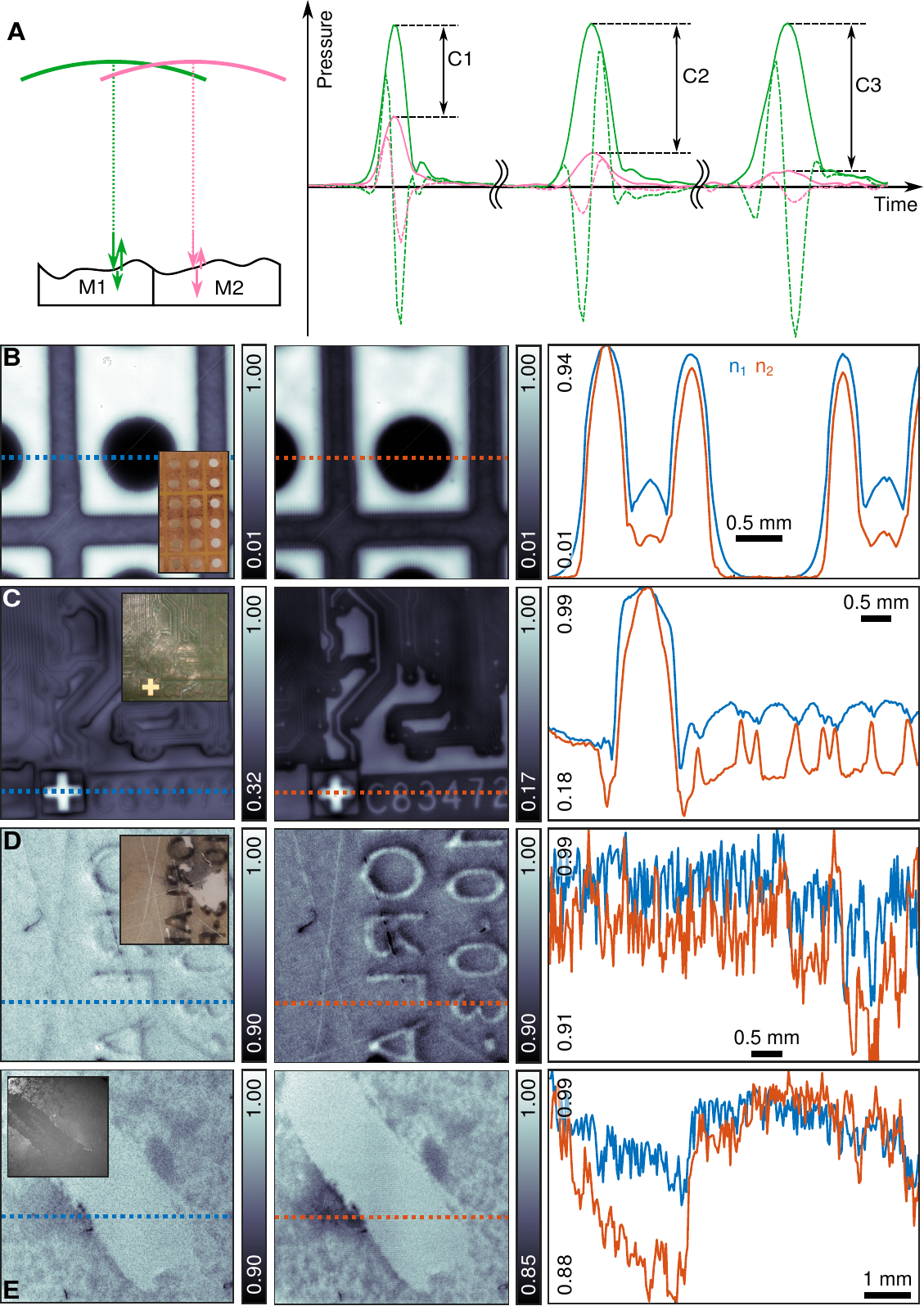}
	\caption{Contrast enhancement with \ac{HOPE} ultrasound. (A) \ac{US} transducer interrogates two different materials (M1, M2). The contrast enhancement (C1, C2, and C3) follows a geometric sequence with the high order reflections that are generated via partial reflections at the interface between coupling medium and the sample. The HOPE images were acquired from (B) Multilayer material consisting of copper and resin-bonded paper, (C) Memory controller microchip, (D) Optical mirror with thin scratches, and (E) \ac{HEK} cells growing on a Petri dish.  The first column in (B)-(E) corresponds to the \ac{MAP} of the first reflection order images with an inset showing a color photograph of each sample. Second reflection orders are shown in the second column, whereas the third column represents the axial signal profiles along the lines depicted in the corresponding MAP images.}
	\label{fig3}
\end{figure}

Contrast in \ac{US} imaging is mainly driven by differences in reflectivity between acoustic boundaries. If the transmitted waveform is reflected multiple times between the boundary and transducer, the resulting signal amplitude will comprise of the material's reflectivity according to a geometric sequence in spatio-temporal reciprocal space. As a result, with increasing number of high order reflection, the information content and contrast between two different reflecting boundaries is expected to increase exponentially in discretized steps (Fig.~\ref{fig3}A).

To showcase the phenomenon, we imaged a prototype circuit board consisting of thin copper layers (approximately \SI{35}{\micro\meter} thickness) placed on top of synthetic resin bonded paper. Due to the higher acoustic impedance of the copper layer (Fig.~\ref{fig3}B), a higher signal amplitude was recorded compared to the signal from the paper or the holes. Next, we analyzed the circuit board of a computer processor (Fig.~\ref{fig3}C) having the copper layers situated beneath a thin layer of laminate. Similar to the previous experiment, we could resolve the embedded circuits with the \ac{CNR} significantly boosted due to diminished signal intensity from the background. \ac{HOPE} was also able to resolve tiny defects on an optical mirror generated with a diamond blade (Fig.~\ref{fig3}D). While some of the scratches are not visible in the first order reflection, they are clearly revealed at the higher order reflections due to the boosted contrast. Features readily visible in the first order reflection appear with increased contrast in the higher order images.

Finally, a single layer of \ac{HEK} cells cultured in a glass Petri dish was imaged with \ac{HOPE}. The cells grow on a flat glass surface representing micron-sized acoustic scatterers due to the acoustic impedance mismatch between cellular structures and the surrounding coupling medium. The scattering effect of the cells increases with each reflection order, thus amplifying contrast between areas in the dish containing the cells versus other areas where the cells were removed prior to imaging (Fig.~\ref{fig3}E). In this case, a contrast enhancement of \SI{7.75}{\decibel} was achieved with \ac{HOPE}.

\paragraph{Self-validation allows for an accurate estimation of frequency-dependent attenuation}

Several key biomedical applications of \ac{US} rely on measurements of acoustic attenuation through biological tissues. Examples include the fabrication of realistic \ac{US} phantoms, assessment of thermal effects of transcranial \ac{US}, or monitoring the disease progression \cite{culjat_review_2010,pinton_attenuation_2011,shishikura_progression_2020,tada_usefulness_2019}. In all those cases, it is paramount to accurately and reliably measure the acoustic attenuation properties of the biological tissues of interest \cite{goss_ultrasonic_1979}. For this purpose, pulse-echo measurements are often performed with a strong reflector placed beneath the tissue sample of a given thickness \cite{okawai_approach_2001,hozumi_ultrasonic_2003,taggart_ultrasonic_2007,arakawa_robust_2018}. In order to calculate the frequency-dependent attenuation properties, a reference measurement is performed where the wavefront travels through the coupling medium without interacting with the sample. Then, the same measurement is repeated but using a planar sample of the tissue with known thickness placed on top of the reflector, and its attenuation properties are calculated by comparing the spectrum of the reflected waveform with and without the sample.

\begin{figure}[h!bt]
	\centering
	\includegraphics[width=0.6\columnwidth]{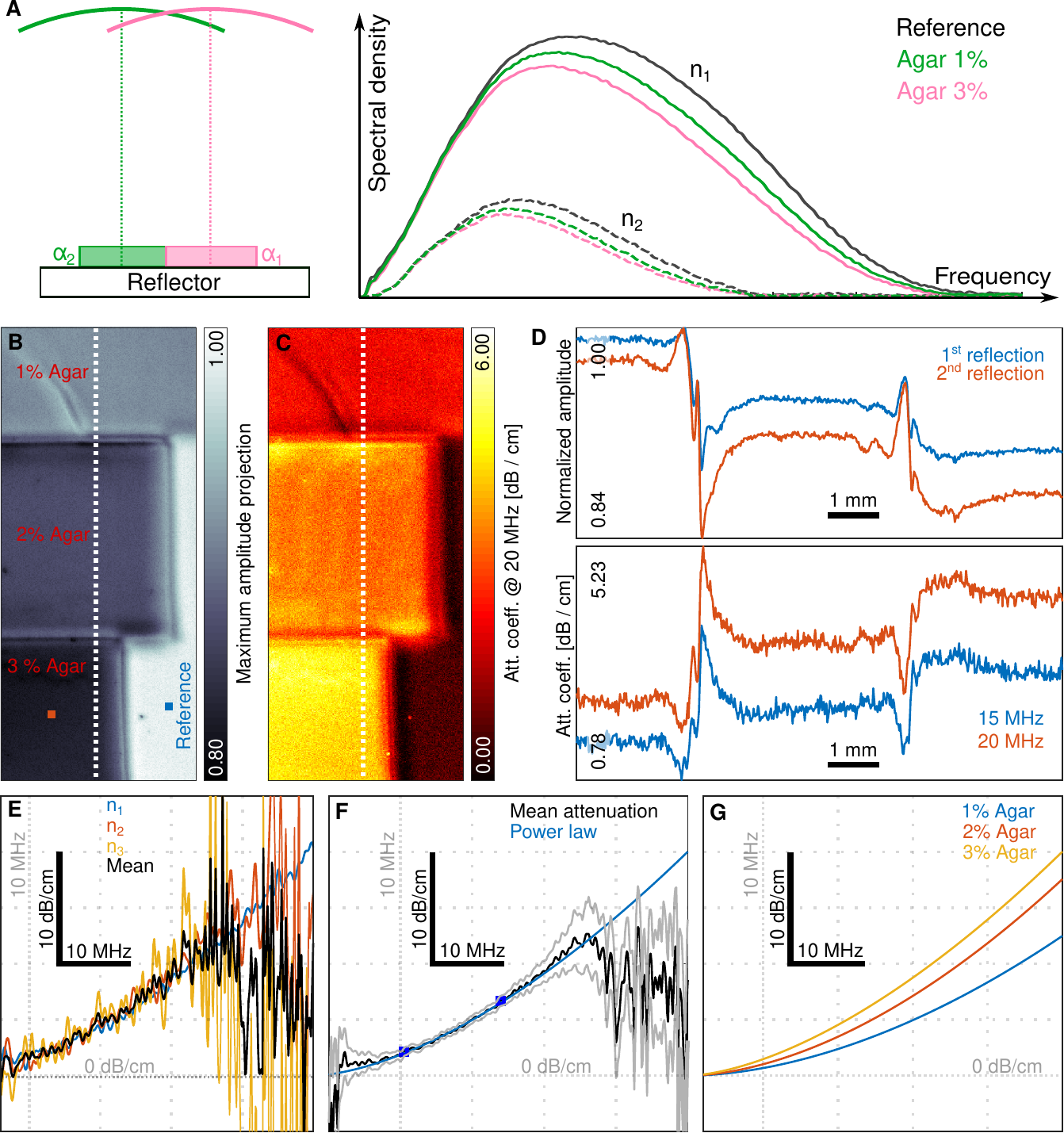}
	\caption{\ac{HOPE} ultrasound attenuation measurement of agar at different concentrations. (A) Self-validated attenuation estimation is based on iterative frequency-dependent measurement taking into account multiple reflections traversing the sample. (B) \ac{MAP} image generated by the first reflection order. (C) Mean estimated attenuation coefficient at \SI{20}{\mega\hertz} calculated from the first three reflection orders. (D) One-dimensional signal profiles for different reflection orders (top) and the corresponding attenuation profiles for different frequencies (bottom). (E) Estimated attenuation coefficient for the different reflection orders at the scanning position marked in red in panel (B). (F) Averaged attenuation coefficient over a 10 x 10 scanning grid surrounding the red mark in (B) with a power law fitted to the data and grey lines showing the standard deviation. (G) Power law fit obtained for the different agar concentrations.}
	\label{fig4}
\end{figure}

Due to the absorption in the coupling medium, the exponential attenuation of US waves, and partial reflection at sample boundary and transducer surface, the initial signal amplitude decreases with each high order reflection (Fig.~\ref{fig4}A). Assuming a constant path of the \ac{US} wave for the different reflection orders, the relative frequency-dependent decline in the signal amplitude between echo $n$ and $n-1$ should remain constant (see proof in Supp. Material). Higher order reflections therefore allow for a self-calibrated mapping of tissue attenuation by comparing ratios extracted from the different reflection orders, which can help distinguishing true acoustic attenuation (absorption) from other effects such as specular reflections at tilted angles. Measuring multiple reflections further allows boosting the measurement’s accuracy with the only setback being reduction in the effective frequency bandwidth (Fig.~\ref{fig4}A).

To validate the proposed procedure, we first measured the attenuation coefficient of \SI{1}{\milli\meter} thin agar slices with increasing concentrations, resulting in increasing attenuation levels. For each scanning position, the attenuation coefficient is estimated using the reference spectrum method for the first three reflection orders, with the mean attenuation coefficient subsequently calculated (Fig.~\ref{fig4}E). The mean attenuation curves obtained at each position are then averaged over a 10 x 10 grid to increase the measurement accuracy and fitted to a power law (Fig.~\ref{fig4}F). This procedure is repeated to obtain the attenuation curves for all agar concentrations (Fig.~\ref{fig4}G), resulting in power law fits with the exponents values of 1.82, 1.77, and 1.61 for the \SI{1}{\percent}, \SI{2}{\percent}, and \SI{3}{\percent} concentrations, respectively. For instance, at frequencies of \SI{15}{\mega\hertz} and \SI{20}{\mega\hertz}, attenuation coefficients of \SI{2.85}{\decibel\per\centi\meter} and \SI{4.55}{\decibel\per\centi\meter} were calculated respectively for the \SI{3}{\percent} agar concentration, which is in excellent agreement with the values reported in literature thus showcasing the high sensitivity of the \ac{HOPE} approach \cite{rabell-montiel_attenuation_2018}. Note that for the \SI{3}{\percent} concentration, the central frequency of the waveform moves away from the initially measured \SI{28.5}{\mega\hertz} frequency to \SI{21.4}{\mega\hertz} and \SI{17.5}{\mega\hertz} for the second and third reflections, respectively.

\begin{figure}[h!bt]
	\centering
	\includegraphics[width=\columnwidth]{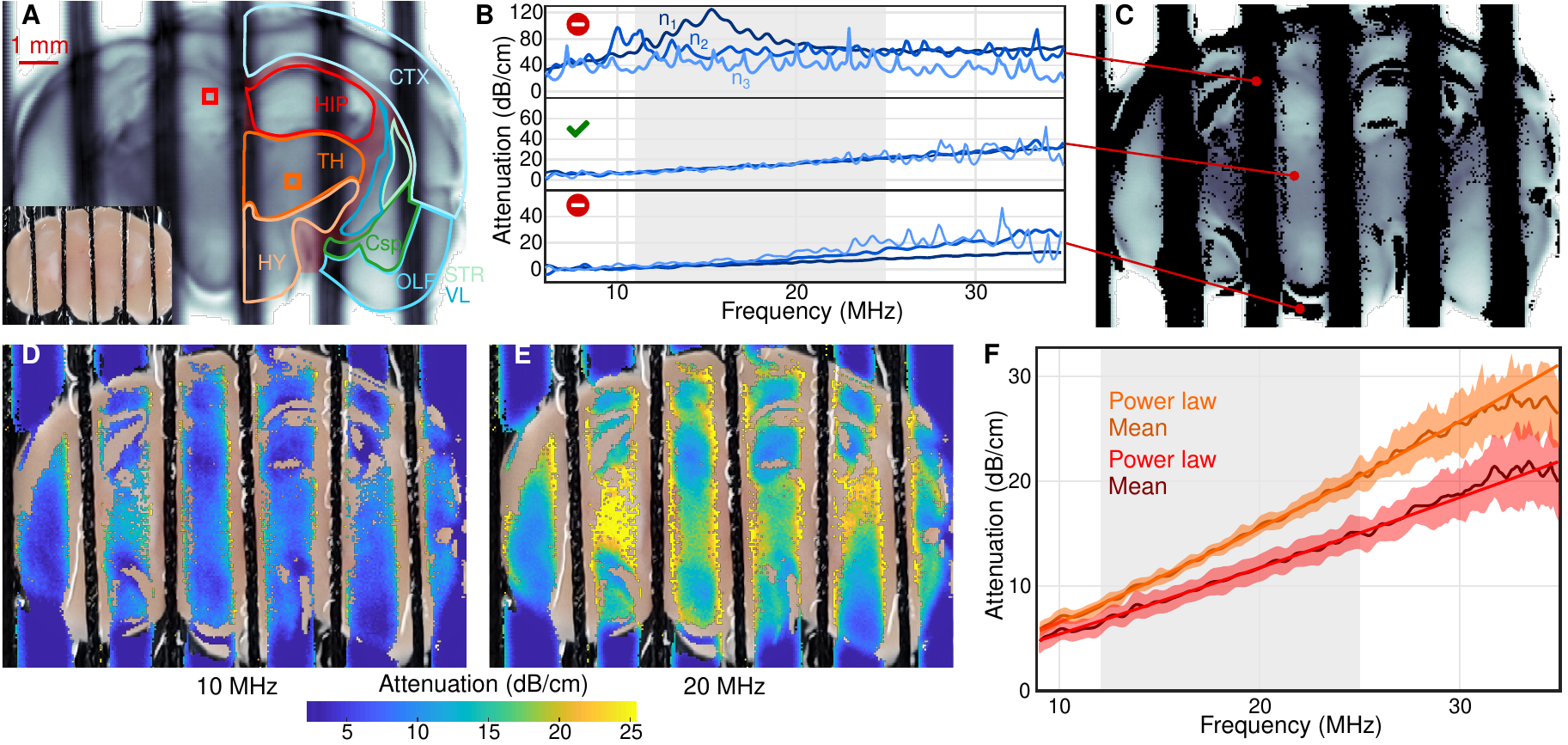}
	\caption{Self-validated \ac{US} attenuation measurement in a mouse brain slice. (A) Maximum amplitude projection of the conventional pulse-echo image. The inset shows the mouse brain and the vertical bars used to keep it fixed during the experiments. Contours delimit the approximate brain regions. (B) Attenuation for reflection orders 1 to 3 in different regions. A binary mask is generated by imposing a \SI{2}{\decibel\per\centi\meter} threshold on deviations among the different reflection orders, defined as the mean absolute error in the gray shaded region. (C). Black regions correspond to points where the measurement is not trusted. Estimated attenuation map at \SI{10}{\mega\hertz} (D) and \SI{20}{\mega\hertz} (E) overlayed on the optical microscope image of the mouse brain. (F) Frequency dependence of the averaged attenuation coefficients over two regions of interests marked in (A) and their corresponding power-law fits. CTX: Cortex, HIP: Hippocampus, TH: Thalamus, HY: Hypothalamus, STR: Striatum, VL: lateral ventricle, OLF: Olfactory area, Csp: Cortical subplate.}
	\label{fig5}
\end{figure}

To demonstrate the self-validation capabilities of \ac{HOPE} ultrasound, we measured the frequency-dependent attenuation of \ac{US} waves traversing a \SI{1}{\milli\meter} thick fresh coronal mouse brain slice. To prevent motion artifacts, the slice was fixed in place using a 3D printed holder with \SI{100}{\micro\meter} Nylon threads. The brain was then immersed in \ac{PBS} at \SI{36}{\celsius} and scanned over a \SI{12}{\milli\meter} x \SI{10}{\milli\meter} region with \SI{50}{\micro\meter} step size (Fig.~\ref{fig5}A). Attenuation was estimated using up to the third reflection order (Fig.~\ref{fig5}B) with three different scenarios identified by the measurement: a) the attenuation diverges due to the presence of a strong scatterer (nylon thread), b) the attenuation converges, and c) the attenuation slightly diverges due to edge reflections. By choosing a divergence threshold, a binary mask can be created to segment trusted versus untrusted regions (Fig.~\ref{fig5}C).  Fiber tracts such as the corpus callosum and alveolus surrounding the hippocampal region (HIP) are segmented as untrusted due to their highly scattering properties. Also, the interfaces between fiber tracts and gray or white matter generate edge reflections. The striatum and adjacent fiber tracts show the highest attenuation (Fig.~\ref{fig5}D, Fig.~\ref{fig5}F). We next compared the attenuation as a function of frequency for a small region of interest from the left HIP and the right thalamus (TH) (Fig.~\ref{fig5}A, Fig.~\ref{fig5}F), which exhibited a characteristic power law behavior. For the \SI{20}{\mega\hertz} frequency, the attenuation changes from \SI{15.42}{\decibel\per\centi\meter} in the TH to \SI{11.69}{\decibel\per\centi\meter} in the HIP. 

\section{Conclusion}

We developed a novel imaging modality taking advantage of high order \ac{US} reflections, previously considered undesired reverberation artifacts \cite{hassler_apparatus_nodate}. The information contained in the high order echoes results in improved image contrast and depth resolution for profilometric measurements. The depth resolution improvement is a consequence of accumulating propagation delays at the high order reflections, which allows to resolve finer structures without increasing the sampling frequency of the data acquisition system. We experimentally demonstrated a four-fold improvement in depth resolution capacity by revealing \SI{745}{\nano\meter}-sized features using a transducer with only \SI{30}{\mega\hertz} bandwidth. On the other hand, the contrast enhancement of high reflection orders allows to visualize defects in materials which are invisible with conventional \ac{SAM} (Fig.~\ref{fig3}D). Such contrast enhancement is chiefly attributed to the iterative interaction between the \ac{US} wavefront and the target, leading to an accumulation of the reflected amplitude produced by two materials with different acoustic properties.

Moreover, \ac{HOPE} ultrasound can be used to improve \ac{US} attenuation measurements. It was demonstrated that for materials having acoustic properties similar to the coupling medium, the attenuation obtained at each reflection order should be identical. If the spatial coherence of the wave deviates from the original assumptions due to specular reflections at tilted angles or other scattering mechanisms, the attenuation for the different reflection orders will diverge. A simple threshold for tolerated divergence will produce a binary mask enclosing the sample regions where the measurement can be trusted. We anticipate that this new self-validated method would help reducing the large variability found in attenuation measurements of biological tissues \cite{culjat_review_2010,goss_ultrasonic_1979}.

Absorption in the coupling medium may diminish the effective bandwidth of the attenuation measurement with respect to conventional pulse-echo measurements done with a single reflection, especially if the system is operated at frequencies beyond \SI{100}{\mega\hertz} where \ac{US} waves get significantly attenuated in water (Fig.~\ref{fig4}A). However, for the reported agar and brain samples, the frequency range of the self-validated attenuation measurement (between \SI{12}{\mega\hertz} and \SI{25}{\mega\hertz}) still provided a useful range to reliably fit a power-law function.

Although \ac{PVDF} polymer provides an excellent platform to sensitive ultrawideband detection of \ac{US} waves, the backing material usually consists of Epoxy resin having limited reflectivity due to its diminished acoustic impedance mismatch with water. New designs might be considered to maximize reflections at the transducer surface thus improving the \ac{SNR} of \ac{HOPE} ultrasound.

High frequency transducers employing a lens made from a solid material could present additional challenges related to reverberations in the lens. Since those are convolved with the target's response for every new reflection, recursive deconvolution approaches would need to be implemented to circumvent this issue and utilize \ac{HOPE} at frequencies exceeding \SI{50}{\mega\hertz}, where the use of solid lenses for beam shaping purposes is often inevitable.

The limitations of the proposed approach are mainly determined by the SNR considerations at high reflection orders as well as the \ac{US} attenuation in the coupling medium. As a result, higher reflection orders can be exploited for improving the image quality as long as location of peaks in the reflected signals can be accurately estimated while the difference in contrast remains above the noise level. \ac{HOPE} ultrasound produces datasets considerably larger than its conventional counterpart, resulting in more extensive memory consumption and computational burden. Yet, the method does not require additional hardware and is easy to implement with existing \ac{SAM} systems, underscoring \ac{HOPE}'s potential to boost imaging performance across a broad range of biomedical imaging and non-destructive testing applications.

Although the examples demonstrated here have immediate application in \ac{US} imaging, considerable scope exists to expand the paradigm shift introduced by \ac{HOPE} into a wider range of mechanical wave regimens. Multiple reflections between the seafloor and water surface are known to affect sonar-based explorations \cite{moura_sonar_1993}. Submarine topography could thus benefit from super-resolved seafloor images with improved contrast \cite{sonogashira_high-resolution_2020}. Multiple reflections further exist on much smaller scales, such as GHz-range surface acoustic waves in solids, where new applications and developments are also anticipated to take advantage of the \ac{HOPE} concept \cite{gustafsson_local_2012}.

\section{Experimental section}

\paragraph{Imaging System} 
A spherically focused \ac{US} transducer (\ac{PVDF}, Precision Acoustics, United Kingdom) was pulsed by a pulser-receiver unit (5073PR, Olympus, USA). The central frequency, bandwidth, radius of curvature, and aperture diameter of the transducer were \SI{30}{\mega\hertz}, \SI{100}{\percent}, \SI{7}{\milli\meter}, and \SI{6.8}{\milli\meter}, respectively. The lateral resolution of the system is limited by diffraction, while the axial resolution is limited by the pulse duration, determined by the transducer bandwidth. Within the focal plane, resolution is diffraction limited to \SI{22}{\micro\meter} and \SI{52}{\micro\meter} in axial and lateral directions, respectively \cite{passmann_vivo_1995}.

Two perpendicularly arranged stages (M-683, PI, Germany and DDSM50/M, Thorlabs, US) scan the transducer in two dimensions along the sample surface, thereby generating three-dimensional images (C-scans). To speed up image acquisition, an overfly scan is implemented relying on a position readout of the fast-moving stage encoder with a microcontroller (Teensy 3.6, PJCR, US). After each incremental step size, both pulser and data acquisition card are triggered. With the described scanning patterns, field of view spanning \SI{5}{\milli\meter} x \SI{5}{\milli\meter} can be acquired within \SI{2}{\minute} at \SI{20}{\micro\meter} lateral resolution \cite{hofmann_rapid_2020}.

The pulser-receiver amplified the received signal by \SI{27}{\decibel} and a data acquisition card digitized the bipolar signal with a resolution of 16 bit, \SI{250}{\mega\hertz} sampling frequency, and ±\SI{5}{\volt} measurement range (M4i.4420, Spectrum Systementwicklung Microelectronic, Germany). A PC running Windows 10 recorded and processed the data (64 Gb RAM, Intel i7 10710U, Intel NUC, USA). Data acquisition was performed with MATLAB R2019b (The MathWorks, USA).

\paragraph{Data preprocessing}
Within each A scan, a negative peak within the first microsecond corresponds to the pulsing event of the pulser-receiver and was interpreted as the zero timepoint $t$ = \SI{0}{\second}. Each A-scan is bandpass filtered between \SI{1}{\mega\hertz} and \SI{80}{\mega\hertz} using a zero-phase 3rd order Butterworth filter to reduce noise. Then, the different peaks of the A-scans are sorted and classified according to the reflection order, and their envelope is calculated using the absolute of a Hilbert transform. All pre-processing and processing calculations were performed using commercial MATLAB R2019b software (MathWorks, USA).

\paragraph{Estimation of the attenuation coefficient}
The attenuation coefficient was calculated at each scanning position for every reflection order using the ratio between the spectrum of the reflection from the sample-glass boundary and the spectrum of the reflected reference signal from the glass reflector without the sample (for more information, see Supplementary Text), that is,

\begin{equation}
	\alpha_S^n(f) \approx \alpha_0(f) - \frac{1}{2 \cdot d \cdot n} \cdot \ln \displaystyle\left\lvert \frac{Y_S^n(f)}{Y_{ref}^n(f)} \right\rvert
\end{equation}

where $\alpha_0(f)$ is the attenuation coefficient of deionized water measured in Np/m, $d$ is the sample thickness, $n$ is the reflection order, and $Y_S^n(f)$ and $Y_{ref}^n(f)$ are the received and the reference spectrum of the n-th reflection order. The attenuation coefficient of distilled water is modelled as a quadratic function $\alpha_0(f) = A(T) f^2$, where $A(T)$ is a 7th order polynomial given as a function of the temperature \cite{treeby_measurement_2011,pinkerton_absorption_1949}.

The estimation is only trusted if good agreement exists among the different reflection orders at the given scanning position, otherwise the estimation is discarded as other effects besides acoustic attenuation may introduce errors. Hence, the attenuation coefficient at each scanning position was calculated as the mean value between estimations obtained at the different reflection orders. Finally, the attenuation coefficient of the three Agar concentration mixtures was obtained by averaging the estimations over uniform regions of interest. A power law fit $\alpha_s(f) = a \cdot f^b$ of the form  was then calculated using the least squares method, being $a$ and $b$ the fitting parameters.

\paragraph{Phantoms}
The prototyping board used for the experiments consists of a layer of synthetic resin bonded paper of \SI{1.5}{\milli\meter} thickness with a \SI{35}{\micro\meter} copper layer glued on top. The employed microchip phantom is a memory controller QG82945G from Intel with a total imaged size of \SI{6}{\milli\meter} x \SI{6}{\milli\meter}. The phantoms employed for the attenuation estimation measurements were prepared using Agar-agar powder for microbiology (Agar-agar, Sigma-Aldrich, USA) and deionized water. Three different agar concentrations were used, namely \SI{1}{\percent}, \SI{2}{\percent}, and \SI{3}{\percent}. The agar-deionized water mixtures were left one day at room temperature to ensure that the mixtures completely solidified, and then cut into \SI{1}{\milli\meter} thin slices. 

\paragraph{Cell cultures}
HEK cells were cultured in Petri dishes having a glass plate at the bottom and incubated for \SI{48}{\hour}. Prior to imaging, the culturing medium was diluted with prewarmed PBS to provide an optimal acoustic coupling for the imaging procedure. Photographs of the samples were acquired prior and post imaging with a USB camera (UI-2340CP-M-GL-TL, Thorlabs, USA) attached to a binocular (Stemi 305, Zeiss, USA). To showcase the contrast between regions with a cell layer versus regions containing no cell layer, a stripe was carefully removed using a Q-tip.

\paragraph{Brain slices}
One C57BL/6J mouse (Janvier, France) was euthanized by ketamin / xylazine overdose. All experiments were performed in accordance with the Swiss Federal Act on Animal Protection and approved by the Cantonal Veterinary Office Zurich. The brain was extracted and cut in \SI{1}{\milli\meter} thick slices. A custom brain-slice holder was designed to prevent motion during the scanning procedure. The slice corresponds to the region around the p56 coronal section (© 2004 Allen Institute for Brain Science. Allen Mouse Brain Atlas, Available at: \href{https://atlas.brain-map.org/}{\texttt{https://atlas.brain-map.org/}}). Photographs of the brain slice were acquired with a USB camera (UI-2340CP-M-GL-TL, Thorlabs, USA) attached to a binocular (Stemi 305, Zeiss, USA).

	\section*{Acknowledgements}
	
	We thank Michael Reiss for animal handling and brain slice preparation, Karolina Werynska for help with the brain slice handling and the help of Beau Le Roy with preparing the HEK cell cultures.
	
	S.P.-L acknowledges financial support from Universitat Politècnica de València research grant PAID-01-18.
	
	UH conceived the idea, developed the theory and the custom-made scanning acoustic microscope, performed preliminary experiments, and assisted the final experiments. UH and SP wrote the manu- script. SP performed the final experiments, evaluated datasets, and developed and performed simulations. UH and HE planned the study. HE conceived the self-validated attenuation measurement, implemented by SP. HE and DR supervised the work, revised the manuscript, and coordinated the work. DR acquired funding for the study. All authors read and commented on the manuscript.

	The authors declare no competing interests.
	
	All data presented in the manuscript is available upon reasonable request. Code for hardware controlling, data acquisition, and postprocessing is available on \href{https://github.com/razanskylab}{\texttt{https://github.com/razanskylab}}.

	\bibliography{library.bib}


\section*{Supporting Information}

\paragraph{Increase in depth resolution}
In general, for a single acoustic boundary located at $z_0$, the $n$-th echo between transducer and target travels a distance given by $z(n) = 2 \cdot n \cdot z_0$. Hence, each time point in an A-scan can be converted into axial distance via
\begin{equation}
	z_i = \frac{c_0 \cdot t_i}{2 \cdot n}
\end{equation}

The minimum difference in axial distance detectable with the system is thus given by the minimum delay or time difference discernable by the system, that is, the sampling time of the analog-to-digital converter. Substituting $t_i = t_s = 1 / f_s$ into Equation (S1) directly yields the depth resolution as a function of the reflection order, as described in Equation (1) of the manuscript. This further implies that the depth resolution of a microscopy system is doubled at the second reflection order compared to the resolution attained by the first reflection order, increased by a factor of three for the third order and so forth.

\paragraph{Contrast enhancement}
Contrast in \ac{US} microscopy is defined as the relative difference between the foreground or target signal, and the background signal, that is,
\begin{equation}
	C(f) = \frac{\lvert Y_f(f) \rvert - \lvert Y_b(f) \rvert}{\lvert Y_b(f) \rvert} = \frac{\lvert Y_f(f) \rvert}{\lvert Y_b(f) \rvert} - 1
\end{equation}

Under the linear regime assumption, the pressure received at the transducer for the n-th reflection order generated by an acoustic boundary located at $z_i$, $Y^n(f)$, can be expressed in the frequency domain via 
\begin{equation}
	Y^n(f) = X_0(f) \cdot R(f)^n \cdot \eta(f)^{n-1} \cdot \exp{(-2~\alpha_0(f)~n~z_i)} \cdot \exp{(-2~j~k_0(f)~n~z_i)}
\end{equation}

where $X_0(f)$ represents the initial transmitted waveform spectrum, $R(f)$ is the reflectance of the target, and $\eta(f)$ is the reflectance of the transducer due to impedance mismatch between the piezoelectric material and the medium. The first exponential term of the equation corresponds to frequency-dependent attenuation in the coupling medium, while the second exponential term corresponds to propagation in the medium, being $k_0(f) = 2~\pi~f / c_0$ the propagation constant. Hence, the contrast of the image can be expressed as a function of the reflection order as

\begin{equation}
	C^n(f) = \displaystyle\left\lvert \frac{R_f(f)}{R_b(f)} \right\rvert^n - 1
\end{equation}

being $R_f(f)$ the reflectance response of the target and $R_b(f)$ the reflectance response of the background. This entails the contrast of the image being multiplied by the power of the reflection order.

\begin{figure}[hbt]
	\centering
	\includegraphics[width=0.8\columnwidth]{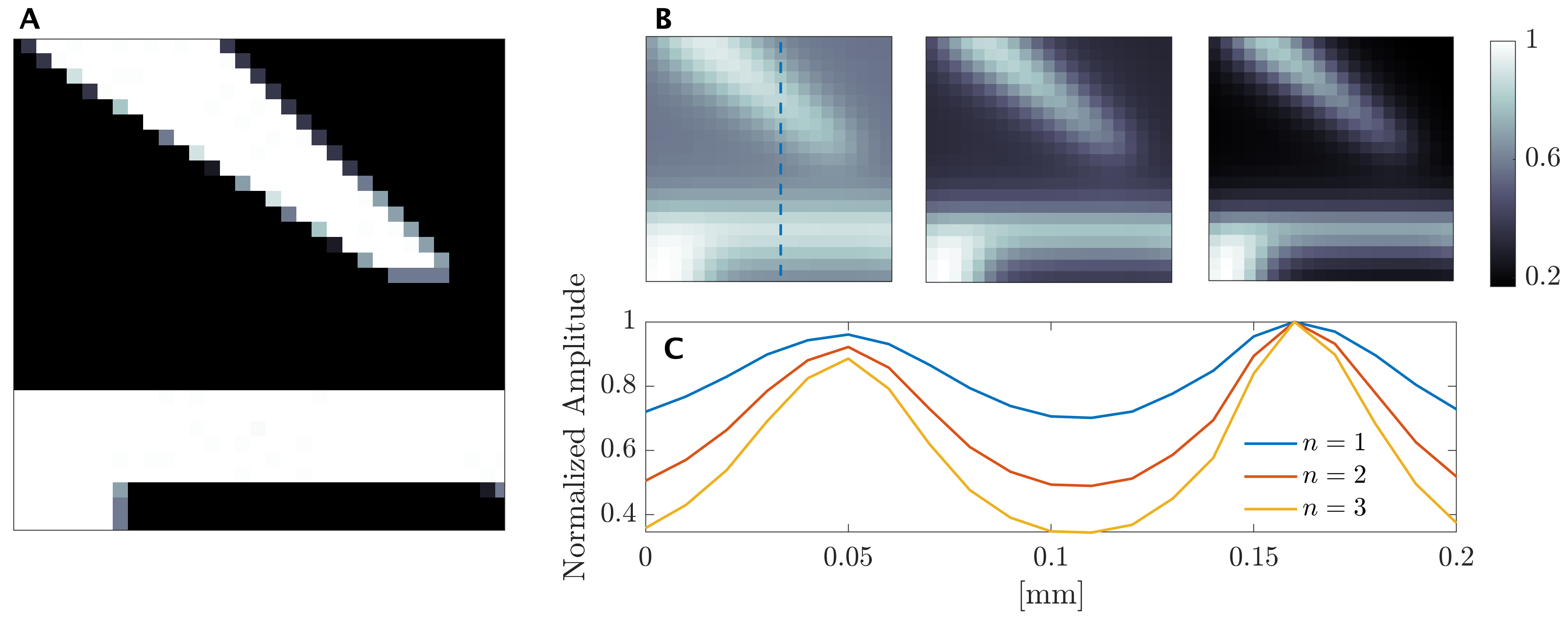}
	\caption{Ideal simulations showcasing the contrast enhancement at higher reflection orders. (A) Initial 2D reflection coefficient distribution, where white regions correspond to a reflection coefficient of  and black regions correspond to . (B) Maximum Amplitude Projections of the first three reflection orders. (C) Axial cuts showing that the amplitude difference between foreground and background increases with the reflection order, and therefore the contrast is enhanced.}
	\label{figs1}
\end{figure}

To showcase this effect, numerical simulations based on the Rayleigh-Sommerfeld integral were performed.[34] For this, we considered a perfectly flat surface having uniform frequency-independent reflectivity located at the focal distance of the transducer. Its parameters were assumed to be the same as for the US PVDF spherically focused transducer used in the experimental measurements.  The pressure distribution at the focal plane of the transducer is then calculated by computing the Rayleigh-Sommerfeld diffraction integral for each simulated frequency, assuming a uniform pressure distribution over the transducer surface. Thereafter, the reflected pressure is calculated by multiplying the ideal reflection coefficient distribution with the incident pressure field. Then, the reflected field is propagated back to the transducer surface using again the Rayleigh-Sommerfeld integral, and the received pressure is calculated as the average pressure over the transducer surface. High order reflections are simulated by reflecting the pressure at the transducer surface back onto the focal plane, repeating the two propagation steps using the Rayleigh-Sommerfeld integral. As can be observed from Fig.~\ref{figs1}, simulations manifest contrast enhancement analogous to that observed in experimental measurements depicted in Fig.~\ref{fig3}. For high order reflections, the contrast is therefore increased by the power of the reflection order.

\paragraph{Attenuation estimation}
The attenuation coefficient is estimated using the relative spectrum method \cite{okawai_approach_2001,hozumi_ultrasonic_2003,taggart_ultrasonic_2007,arakawa_robust_2018}. In this technique, a strong reflector is placed at the focal distance of the transducer and calibrated to be as flat as possible. Then, the soft target sample with a known thickness $d$, i.e. biological or non-biological tissues with acoustic impedance similar to that of the coupling medium, is placed on top of the reflector. For the first reflection order, the signal measured from the reflector located at $z_0$, which is used as a reference, and can be formulated as

\begin{align}
Y_{ref}^1(f) & = X_0(f) \cdot R_{13}(f) \cdot \exp{(-2 (\alpha_0(f) + j k_0(f)) \cdot (z_0 - d))} \cdot \dots \\
	\dots & \exp{(-2 (\alpha_0(f) + j k_0(f)) \cdot d)}
\end{align}

where the first and the second exponential terms represent propagation and attenuation at the coupling medium up to the sample and within the sample axial range, respectively, and $R_{13}(f)$ represents the reflection coefficient of the coupling medium and reflector interface. On the other hand, the signal measured from the sample consists mainly of the sum of two pressure contributions: one reflection at the coupling medium-sample boundary, and the other reflected at the sample-reflector boundary, which can be described, respectively, as

\begin{equation}
	Y_s^1(f) = X_0(f) \cdot R_{12}(f) \cdot \exp{(-2 (\alpha_0(f) + j k_0(f)) \cdot (z_0 - d))}
\end{equation}

\begin{align}
	Y_B^1(f) = & X_0(f) \cdot T_{12}(f) \cdot R_{23}(f) \cdot T_{21}(f) \cdot \dots \\
	\dots & \exp{(-2( \alpha_0(f) + j k_{0}(f)) \cdot (z_{0} - d))} \cdot \dots \\
	\dots & \exp{(-2( \alpha_S(f) + j k_{s}(f)) \cdot d)}
\end{align}

being $T_{ij}(f)$ and $R_{ij}(f)$ the transmission and reflection coefficients of the acoustic boundary between medium $i$ and medium $j$, and $\alpha_s(f)$ and $k_s(f)$ the attenuation and propagation coefficients of the sample. Here, two simplifications can be made. First, if the thickness of the sample is thick enough, it is possible to process each reflection independently by selecting the proper time window. Second, if the acoustic impedance of the sample is very close to that of the coupling medium, which holds true in case of soft biological tissues, the reflection at the boundary between coupling medium and sample can be neglected, as $R_{12}(f)$ would be very close to 0. Therefore, reflection from the sample can be estimated by only considering the sample-reflector interface. With this in mind, it is possible to define a relative amplitude spectrum as the ratio between the spectrum of the sample reflection and the reference reflection from the reflector, that is,

\begin{equation}
	\displaystyle\left\lvert \frac{Y_B^1(f)}{Y_{ref}^1(f)} \right\rvert = 
	\displaystyle\left\lvert \frac{T_{12}(f) R_{23}(f) T_{21}(f)}{R_{13}(f)} \right\rvert \cdot
	\displaystyle\left\lvert \frac{\exp{(-2 \alpha_s(f) d)}}{\exp{(-2 \alpha_0(f) d)}} \right\rvert
\end{equation}

The above equation can be simplified by taking logarithms, leading to an estimated expression for the sample attenuation using the first order reflection, i.e. 

\begin{equation}
	\alpha_s^1(f) = \frac{1}{2 d} \cdot \ln{\displaystyle\left\lvert \frac{T_{12}(f) R_{23}(f) T_{21}(f)}{R_{13}(f)} \right\rvert} + \alpha_0(f) 
		- \frac{1}{2 d} \cdot \ln{\displaystyle\left\lvert \frac{Y_B^1(f)}{Y_{ref}^1(f)} \right\rvert}
\end{equation}

where the attenuation in the coupling medium $\alpha_0(f)$, in our case deionized water, is a known variable modelled as a quadratic frequency function with a temperature-dependent amplitude factor [32, 33]. If the approximations described above can be made, which, as previously stated, is the case for biological tissues, the above equation can further be simplified since $T_{12}(f)$ and $T_{21}(f)$ would be approximately 1. Thus, 

\begin{equation}
	\alpha_s^1(f) \approx \alpha_0(f) 
		- \frac{1}{2 d} \cdot \ln{\displaystyle\left\lvert \frac{Y_B^1(f)}{Y_{ref}^1(f)} \right\rvert}
\end{equation}

Analogously, the expression for the attenuation coefficient estimation for the $n$-th order reflection can be obtained via

\begin{equation}
	\alpha_s^n(f) \approx \alpha_0(f) 
		- \frac{1}{2 d} \cdot \ln{\displaystyle\left\lvert \frac{Y_B^n(f)}{Y_{ref}^n(f)} \right\rvert}
\end{equation}

which corresponds to Equation (2) of the Materials and Methods section.
Thus, the estimated attenuation coefficient should be identical for all the reflection orders, provided that the simplified assumptions hold true. This property can be used to test the validity of the assumptions for a given sample. 

\end{document}